# Self-supervised physics-informed generative networks for phase retrieval from a single X-ray hologram

Xiaogang Yang,[1,†,*] Dawit Hailu,[2] Vojtěch Kulvait,[2] Thomas Jentschke,[2] Silja Flenner,[2] Imke Greving,[2] Stuart I. Campbell,[1] Johannes Hagemann,[3] Christian G. Schroer,[3,4,5] Tak Ming Wong,[2,6] and Julian Moosmann[2,†]

[1]*NSLS-II, Brookhaven National Laboratory, Upton, NY, 11973, USA*
[2]*Institute of Materials Physics, Helmholtz-Zentrum Hereon, Max-Planck-Straße 1, 21502 Geesthacht, Germany*
[3]*Center for X-ray and Nano Science CXNS, Deutsches Elektronen-Synchrotron DESY, Notkestraße 85, 22607 Hamburg, Germany*
[4]*Department of Physics, Universität Hamburg, Luruper Chaussee 149, 22761 Hamburg, Germany*
[5]*Helmholtz Imaging, Deutsches Elektronen-Synchrotron DESY, Notkestraße 85, 22607 Hamburg, Germany*
[6]*Institute of Metallic Biomaterials, Helmholtz-Zentrum Hereon, Max-Planck-Straße 1, 21502 Geesthacht, Germany*
[†]*The authors contributed equally to this work.*
[*]*yangxg@bnl.gov*

**Abstract:** X-ray phase contrast imaging significantly improves the visualization of structures with weak or uniform absorption, broadening its applications across a wide range of scientific disciplines. Propagation-based phase contrast is particularly suitable for time- or dose-critical in vivo/in situ/operando (tomography) experiments because it requires only a single intensity measurement. However, the phase information of the wave field is lost during the measurement and must be recovered. Conventional algebraic and iterative methods often rely on specific approximations or boundary conditions that may not be met by many samples or experimental setups. In addition, they require manual tuning of reconstruction parameters by experts, making them less adaptable for complex or variable conditions. Here we present a self-learning approach for solving the inverse problem of phase retrieval in the near-field regime of Fresnel theory using a single intensity measurement (hologram). A physics-informed generative adversarial network is employed to reconstruct both the phase and absorbance of the unpropagated wave field in the sample plane from a single hologram. Unlike most state-of-the-art deep learning approaches for phase retrieval, our approach does not require paired, unpaired, or simulated training data. This significantly broadens the applicability of our approach, as acquiring or generating suitable training data remains a major challenge due to the wide variability in sample types and experimental configurations. The algorithm demonstrates robust and consistent performance across diverse imaging conditions and sample types, delivering quantitative, high-quality reconstructions for both simulated data and experimental datasets acquired at beamline P05 at PETRA III (DESY, Hamburg), operated by Helmholtz-Zentrum Hereon. Furthermore, it enables the simultaneous retrieval of both phase and absorption information.

## 1. Introduction

X-ray phase contrast imaging has become a vital tool for exploring materials with low attenuation contrast, driving progress in fields such as biology, materials science, archaeology, semiconductor technology, energy storage, pharmaceuticals, and food science [1]. Phase contrast techniques provide the sensitivity needed to visualize fine structural details in weakly attenuating samples where conventional X-ray imaging lacks intensity contrast [2]. Various methods for X-ray phase contrast imaging have been developed, such as propagation-based (or in-line) [3–7], analyzer-based [8], interferometric [9,10], aperture-based [11,12] or speckle-based methods [13–16]. Here, we only consider propagation-based phase contrast methods that measure the intensity of the transmitted and forward-propagated X-ray wave front at a single distance behind the sample. As the X-rays traverse the sample, the wave front is attenuated and accumulates a (negative) phase shift. Upon exiting the sample, the transmitted wave front is diffracted and additional intensity contrast emerges through self-interference upon free-space propagation. See Fig. 1 for a propagation-based phase contrast setup at a nano-tomography end station at a synchrotron-radiation facility. Single-distance phase contrast imaging without additional X-ray optics or extensive scanning procedures is ideal for time-sensitive or dose-critical applications such as in situ, operando, or in vivo experiments.

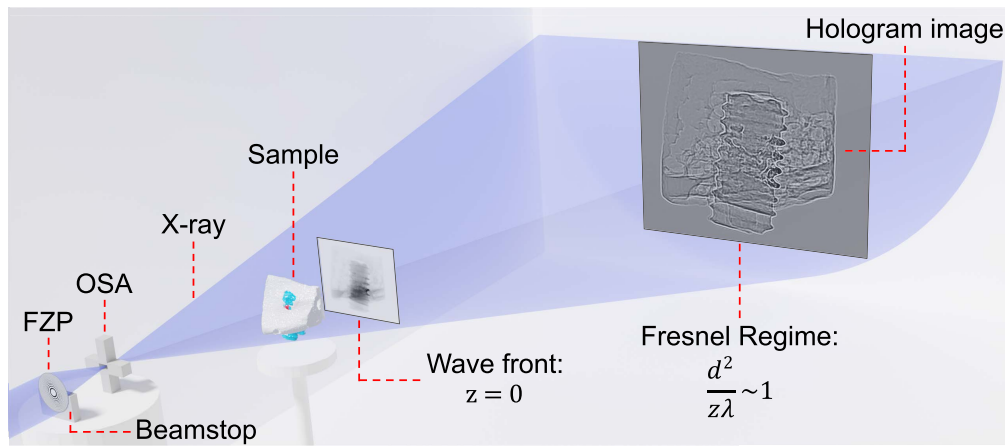

**Fig. 1.** Experimental setup for an in-line X-ray holography measurement at a nano-tomography beamline at a synchrotron-radiation facility. The X-ray beam is generated in an undulator and monochromatized using a double crystal monochromator. The beam is focused using a Fresnel zone plate (FZP). The zero diffraction order is blocked by the beamstop and the higher orders by the order sorting apertures (OSA). X-rays impinge on the sample at propagation distance $z = 0$. During object transmission, X-rays are attenuated and accumulate a phase shift. Upon exiting the object, the transmitted wave front is diffracted. During forward propagation in free space, intensity contrast emerges due to self-interference of the diffracted wave front.

Due to the high-frequency oscillations of X-rays, state-of-the-art high-resolution detector systems cannot measure the phase of the propagated wave front. Only the intensity, i.e., the time-averaged square amplitude of the complex-valued wave field, is accessible. Thus, propagation-based phase contrast imaging requires recovering the phase (and amplitude) information in the object plane from a single intensity measurement before further data analysis. This represents an ill-posed inverse problem. In general, obtaining a unique, physically consistent solution is inherently challenging, as the same measured intensity can arise from multiple combinations of phase and amplitude [17]. The commonly used phase retrieval methods are based on scanning techniques [18–22], multiple intensity measurements at different distances [4,23], or certain approximations and constraints imposed by analytical methods. In the former case, the data acquisition time and dose are considerably increased. In the latter, constraints imposed by analytical methods based, e.g., on the transport of intensity equation (TIE) [24], the contrast transfer function (CTF) [25], or assumptions that the material under test is a pure phase object, solely composed of a single material [26] or that there is a phase-attenuation duality for high X-ray energies [27], significantly limit the applicability of these methods. These limitations can be overcome by using classical iterative schemes such as Gerchberg-Saxton (GS) [28] or hybrid input-output (HIO) algorithms [29,30]. Although described 50 years ago, their analytical properties are still not precisely known [31], and in practice, their convergence strongly depends on the particular specification of boundary conditions and other constraints requiring expert knowledge.

Deep learning has recently emerged as a powerful alternative to traditional iterative methods for solving ill-posed inverse problems in imaging [32]. Among these, phase retrieval has seen growing interest, with machine learning approaches increasingly explored as a means to improve reconstruction quality and efficiency [33,34]. Supervised learning approaches have utilized large datasets of microscopic images [35–39]. For example, ResNet and U-Net architectures were trained on experimentally generated paired data to recover pure-phase images [35,36], while simulated paired data were used in other studies [40]. Classical phase retrieval methods provided ground-truth amplitude and phase images for the training of convolutional neural networks (CNNs) [34,37,38]. Cross-modality networks enabled 3D reconstructions from single in-line holograms [41], and post-processing techniques, such as auto-focusing, further improved results [38]. Neural networks have also served as regularizers within optimization frameworks [42,43]. Supervised and unsupervised physics-informed approaches were employed in Fourier ptychography [44]. Multi-resolution, physics-guided Bayesian CNNs have additionally addressed inverse problems, including uncertainty quantification [45]. These studies showed that deep learning works effectively for phase retrieval, emphasizing the need for high-quality training data. However, acquiring sufficient training data remains challenging, particularly in synchrotron-radiation phase contrast tomography, due to limited facility access and significant variability in sample characteristics and experimental setups. Generating accurate digital phantoms or simulations is often difficult due to uncertainties in sample composition and morphology, which further limit the availability of reliable training datasets.

Beyond data-driven learning, the integration of physics-based models has proven to be effective in solving the inverse problem of Fresnel propagation, reducing the reliance on large datasets while embedding physical constraints. Deep-learning-based regression methods, such as residual-block CNNs, U-Nets, ResNets, and generative adversarial networks (GANs), have improved phase retrieval by mitigating noise and aliasing. Further advancements in pre-processing, composite loss functions, and attention mechanisms have enhanced their accuracy and robustness [46]. The Deep Gauss–Newton (DGN) algorithm is a learned iterative scheme, which is obtained by unrolling a Gauss–Newton iteration [43]. It includes the knowledge of the imaging physics and is trained on simulated, paired image data. PhaseGAN is a GAN-based network that eliminates the need for paired data in phase recovery by integrating the physics of image formation, allowing it to be trained on unpaired experimental or synthetic data [47]. A similar approach was taken with a cycle GAN trained on unpaired experimental data to recover phase-only objects without prior knowledge [48]. AutoPhaseNN is an unsupervised, physics-informed method that utilizes a 3D convolutional encoder-decoder architecture for inverting 3D far-field coherent diffraction data. It is initially trained on small simulated 3D diffraction patterns and later refined using experimental data [49]. Another machine learning-based hologram reconstruction technique, GedankenNet, leverages artificial images and a Fresnel propagator for training. But unlike our approach, it requires multiple object-to-hologram distances [50]. PhysenNet follows a similar physics-based strategy without requiring training data, but it does not account for absorption and is restricted to phase-only wave fields, significantly simplifying the problem [51]. A physics-informed GAN has also been developed that incorporates forward and backward propagation, adaptive background masking, and a smoothness constraint to enhance reconstruction accuracy [52]. These studies demonstrate the effectiveness of advanced neural architectures, physics-informed methods, and unsupervised learning to improve the accuracy and applicability of phase retrieval.

For self-supervised learning, the pioneer work on deep image priors [53] was considered as one of the most prominent methods for learning without massively labeled data. Ulyanov et al. showed that the prior knowledge of inverse image reconstruction problems can be modeled by the randomly initialized weights of neural networks, which served as a parametrization of the desired image and out-performed traditionally hand-crafted priors. Since the proposal of this ground-breaking approach, deep image prior approaches have become one of the dominant trends in self-supervised learning to solve inverse imaging problems. Qayyum et al. later presented a comprehensive review [54] in which they introduced the term untrained neural network priors (UNNP). However, researchers have demonstrated that the deep image prior approach can be stuck in local minima as discussed in [55, Fig. 3] and [56, Fig. 1]. While deep image prior methods have shown potential in phase retrieval [51,57], there is limited evidence showing their practical applications in X-ray measurements due to the inherently ill-posed nature of the inverse problem. In comparison, self-supervised reconstruction approaches based on generative models have demonstrated improved performance when tackling similarly ill-posed inverse problems [58,59].

In this paper, we introduce SelfPhish: a Self-supervised, Physics-Informed generative networks approach for phase retrieval from a Single X-ray Hologram. We propose a self-training scheme for a generative network to minimize the discrepancy between the measured input hologram and the reconstructed hologram, which is obtained applying the forward physics model to the complex wave field predicted by its generator network. We construct generator and discriminator networks, where the generator predicts the phase shift and absorbance (i.e., the negative logarithm of the amplitude) as image outputs, while the discriminator evaluates whether the reconstructed hologram is consistent with the measured data. The proposed GANs are implemented using TensorFlow and PyTorch, enabling efficient testing and deployment on modern GPU-accelerated platforms. We demonstrate that the networks reconstruct the unpropagated wave field, in terms of the phase shift and absorbance at the object exit plane, from simulated single intensity measurements (i.e., the holograms) at the detector plane. The proposed method is evaluated quantitatively and qualitatively using X-ray holography data from simulations and experimental measurements. Additionally, the proposed method allows the reuse of tuned network weights for reconstructions with similar features or configurations, reducing the need to train from scratch. This leads to faster reconstruction and improved efficiency.

## 2. Proposed method

In this section, we describe the major components of the proposed approach, which are the forward physics model, the self-supervised learning approach, and the generative networks.

### 2.1. Fresnel propagation as the forward physics model

In this work, we only consider non-scanning full-field imaging techniques in order to recover the phase information from a single measurement at a single propagation distance. Furthermore, we only consider forward-propagating wave fields according to the paraxial approximation [5]. For the case of a diverging cone beam, we use the Fresnel scaling theorem [5]. Assuming the object to be described by the refractive index

$$n = 1 - \delta + i\beta \tag{1}$$

and employing the projection approximation [5], the absorbance $A$ and phase $\phi$ of the transmitted wave field at the object exit plane at $z = 0$ are given as projections of the refractive index

$$A(x, y) = \frac{2\pi}{\lambda} \int \beta(x, y, z)\, dz,$$
$$\phi(x, y) = -\frac{2\pi}{\lambda} \int \delta(x, y, z)\, dz, \qquad (2)$$

where $\lambda$ denotes the wave length, $(x, y)$ the transversal spatial coordinates, and we have assumed propagation along $z$. The corresponding transmitted scalar wave field $\psi_0$ in the sample plane at $z = 0$ thus reads

$$\psi_0(x, y) = \exp\left[-A(x, y) + i\phi(x, y)\right], \qquad (3)$$

where we have assumed a homogeneous wave field of unit amplitude impinging on the object. We further assume that the planar wave field $\psi_0$ at the object exit plane is paraxially propagated to the detector plane at a distance $z$. The forward propagated wave-field $\psi_z(x, y)$ is described by Fresnel theory, see [5,60],

$$\psi_z(x, y) = P_z[\psi_0](x, y) = \frac{e^{i2\pi z/\lambda}}{i\lambda z} \int_{-\infty}^{+\infty}\int_{-\infty}^{+\infty} \psi_0(x', y') \exp\left\{\frac{i\pi}{\lambda z}\left[(x - x')^2 + (y - y')^2\right]\right\} dx'\, dy', \qquad (4)$$

where $P_z[\cdot]$ denotes Fresnel propagation related to the propagation distance $z$. Employing the Fourier convolution theorem [5], Eq. (4) reads

$$\psi_z(x, y) = \mathcal{F}^{-1}\left\{\exp\left[-\frac{i\lambda z}{4\pi}(k_x^2 + k_y^2)\right]\mathcal{F}\{\psi_0\}\right\}, \qquad (5)$$

where $\mathcal{F}$ and $\mathcal{F}^{-1}$ denote the forward and inverse 2D Fourier transform, and $(k_x, k_y)$ are the coordinates of the reciprocal domain. Given the (effective) pixel size of the detector $d$, the dimensionless Fresnel number is defined as

$$F = \frac{d^2}{\lambda z}. \qquad (6)$$

The Fresnel propagator, and consequently the propagated wave field, depend on the parameter $F$. For simplicity, however, we continue to use the index $z$. Using Eq. (5) to calculate the Fresnel propagation, the sampling rate N in Fourier space must satisfy the following inequality [61]

$$N \geq \frac{1}{F}. \qquad (7)$$

In the following, phase and absorbance were padded so that Eq. (7) is fulfilled.

Due to the high-frequency oscillation of X-rays, only the square amplitude of the complex wave field $\psi_z$ in Eq. (4) can be measured and the phase information is lost. The measured hologram is a discrete real image $I_z \in \mathbb{R}^{n_x \times n_y}$ with $n_x$ and $n_y$ as the number of pixels in $x$ and $y$ direction respectively, representing the measured intensity at the propagation distance $z$:

$$I_z(x, y) = |\psi_z(x, y)|^2. \qquad (8)$$

The objective of phase retrieval from a single hologram is to retrieve the absorbance $A \in \mathbb{R}^{n_x \times n_y}$ and the phase shift $\phi \in \mathbb{R}^{n_x \times n_y}$ of the wave field in the sample plane from the measurement of the intensity $I_z \in \mathbb{R}^{n_x \times n_y}$ at the detector plane. This is an ill-posed problem as it involves the reconstruction of $A$ and $\phi$ from a single measurement.

## 2.2. Principle of self-supervised learning scheme

In this section, we mathematically express the self-supervised learning strategy of SelfPhish based on the physics model in Section 2.1. We specifically compare to supervised training, unsupervised training, the deep image prior approach and the proposed self-supervised learning under the context of phase retrieval in X-ray phase contrast imaging.

Phase retrieval using a single intensity measurement is typically expressed as the optimization problem [20,28,30,62] for estimating the desired variables $\phi$ and $A$ from the measurement $I_z$ at the spatial position $(x, y)$, given a forward model $f$ (i.e., $f = \psi_z$ in Eq. (5) in our case), based on Eq. (3), Eq. (5), and Eq. (8):

$$\arg\min_{\phi} \sum_{x,y} \mathcal{L}\left[|f(\phi, A)|^2, I_z\right] + R(\phi, A), \tag{9}$$

where $\mathcal{L}$ is a discrepancy measure, such as the mean square error (MSE) loss or the cross-entropy loss, and $R$ denotes a regularizer. Supervised training approaches [47,48] train a network-based model $M$ based on its network parameters $\mathbf{w}$, training batches $\mathbf{b}$ and the (ground-truth) labels $(\phi_{gt}, A_{gt})$:

$$\arg\min_{\mathbf{w}} \sum_{\mathbf{b}} \mathcal{L}\left[M(I_z^{\mathbf{b}}; \mathbf{w}), \phi_{gt}^{\mathbf{b}}, A_{gt}^{\mathbf{b}}\right]. \tag{10}$$

Unsupervised training approaches [49] decode the parameterized representation from the desired variables back to the input variables via the forward model $f$, which yields

$$\arg\min_{\mathbf{w}} \sum_{\mathbf{b}} \mathcal{L}\left\{\left|f\left[M(I_z^{\mathbf{b}}; \mathbf{w})\right]\right|^2, I_z^{\mathbf{b}}\right\}, \tag{11}$$

where $I_z^{\mathbf{b}}$ denotes the training batches of intensity measurements. After training the network $M$ to obtain the optimal weights $\mathbf{w}_{\text{opt}}$, inferences of both supervised and unsupervised training approaches pass the measured hologram $I_z$ through the network to get the desired variable, i.e. $(\phi, A) = M(I_z; \mathbf{w}_{\text{opt}})$.

However, the generalization of these approaches relies on the quality and quantity of the training data, which is challenging for X-ray phase contrast experiments. Typically, experimental data with known phase and absorbance is unavailable, and simulated training data is also lacking, as the composition and morphology of the sample are not yet known. Hence, deep image prior approaches can be beneficial for laboratory environments such as synchrotron-radiation facilities where phase contrast imaging experiment are conducted. Deep image prior approaches (such as [53]) typically work on a single image instead of batches, and directly minimize the discrepancy between the predicted parameter (i.e., the network output) and the desired observations (i.e., the input variable):

$$\arg\min_{\mathbf{w}} \sum_{x,y} \mathcal{L}\left\{\left|f[M(I_z; \mathbf{w})]\right|^2, I_z\right\}. \tag{12}$$

Notice that the network model $M$ in Eq. (12) parameterizes the desired variables $(\phi, A)$ in Eq. (9) such that the optimization problems of Eq. (12) and Eq. (9) are similarly minimizing the discrepancy over the entire image.

However, unlike optimization problems such as image denoising and inpainting as shown in [53] which predict the desired image from the observation image $M : \mathbb{R}^{n_x \times n_y} \to \mathbb{R}^{n_x \times n_y}$, phase retrieval methods predict two variables, the phase shift $\phi$ and the absorbance $A$, based on an input hologram $I_z$, which yields a generative problem $M : \mathbb{R}^{n_x \times n_y} \to \mathbb{R}^{n_x \times n_y \times 2}$ for the network.

Therefore, we propose using generative networks, which is inspired by GANrec [58], to train a generator network $G(I_z; \mathbf{w}_G)$ and a discriminator network $D(I_z; \mathbf{w}_D)$ to minimize the discrepancy and to maximize the likelihood, yielding

$$\arg\min_{\mathbf{w}_G} \max_{\mathbf{w}_D} \sum_{x,y} \mathcal{L}\left(D\left\{|f[G(I_z;\mathbf{w}_G)]|^2;\mathbf{w}_D\right\}, D(I_z;\mathbf{w}_D)\right). \tag{13}$$

This indicates that the desired variables are parameterized by the network parameters, and optimizing these parameters serves as a self-supervised learning method for the phase retrieval problem, without requiring any labeled data.

Based on the aforementioned self-supervised learning approach and the Fresnel propagation model Eq. (3) and Eq. (5) described in Sec. 2.1, we formulate an adversarial training framework. We employ the sigmoid cross-entropy loss as the adversarial term, consistent with our self-supervised formulation, and use the $L_1$-loss as a data discrepancy term. The latter not only promotes faster convergence but also ensures physical consistency with the measured intensity data. For the sake of simplicity, as illustrated in Fig. 2, we abbreviate the generator network and its weights with $G$ and the discriminator network with its weights with $D$, i.e., $G : \mathbb{R}^{n_x \times n_y} \to \mathbb{R}^{n_x \times n_y \times 2}$ and $D : \mathbb{R}^{n_x \times n_y} \to [0, 1]$,

$$\begin{aligned}
I_g &= |P_z[G(I_z)]|^2 \\
\mathcal{L}_{\text{GAN}}(I_z; G, D) &= \mathbb{E}\left\{\log S[D(I_z)]\right\} + \mathbb{E}\left(\log\left\{1 - S[D(I_g)]\right\}\right) \\
\mathcal{L}_{\text{data}}(I_z; G) &= \sum \|I_g - I_z\|_1,
\end{aligned} \tag{14}$$

where $P_z$ denotes the Fresnel propagation of Eq. (4), $I_z$ and $I_g$ represent the measured intensity of Eq. (8) and the generated intensity, respectively, and $\mathbb{E}$ and $S$ are the expected value and the sigmoid operators, respectively. We use the adversarial model with objective function Eq. (14) to

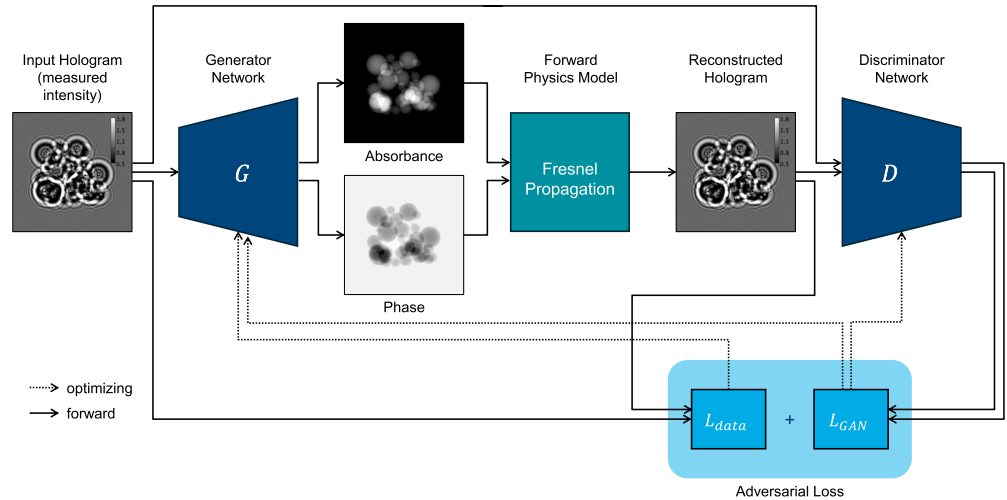

**Fig. 2.** SelfPhish is a self-supervised learning approach based on a generative adversarial network (GAN) and the physics model for the phase retrieval problem using a single intensity measurement, i.e., hologram. The generator network $G$ predicts the desired variables, i.e., absorbance and phase. The forward model is given by Fresnel theory and calculates the hologram based on the predicted absorption and phase. The discriminator network $D$ classifies whether the reconstructed hologram is similar to the measurement hologram.

find the optimal generator network $G^*$ and the corresponding optimal reconstruction $(\phi^*, A^*)$:

$$G^* = \arg\min_G \max_D \mathcal{L}_{\text{GAN}}(I_z; G, D) + \gamma \mathcal{L}_{\text{data}}(I_z; G)$$
$$\psi_z^*(x, y) = \exp(-A^* + i\phi^*) = G^*[I_z(x, y)], \quad (15)$$

where $\gamma$ denotes a coefficient to control the weight of the penalization term

In this section, we provide details of the generative adversarial networks (GANs), including the network architectures, implementation aspects and the optimization of network weights. Additional implementation details for both TensorFlow and PyTorch are available in the source code repositories on GitHub at https://github.com/XYangXRay/selfphish and https://github.com/daveabiy/selfphish, respectively.

**Networks architecture** Our architecture consists of two networks, one for the generator (G) and one for the discriminator (D), as shown in Fig. 3 and Fig. 4 respectively. Inspired by the GANs from GANrec [58], the generator consists of fully connected (FC) and convolution layers. We use convolutional layers to refine the output of the fully connected layers. The convolutional layers require a much smaller number of weights than the fully connected layers. The generator uses a normalized, i.e., flat-field corrected intensity (hologram) as input which is flattened to a 1D array, passed through a sequence of fully connected layers, reshaped to the original input size, and then passed to convolutional and deconvolutional layers. The last layer has two output channels corresponding to phase $\phi_{\mathbf{w}}$ and absorbance $A_{\mathbf{w}}$, respectively, which gives a possible exit wave function according to Eq. (15) which afterwards is propagated to the detector according to Eq. (5). The square amplitude of the propagated wave field is then input for the discriminator.

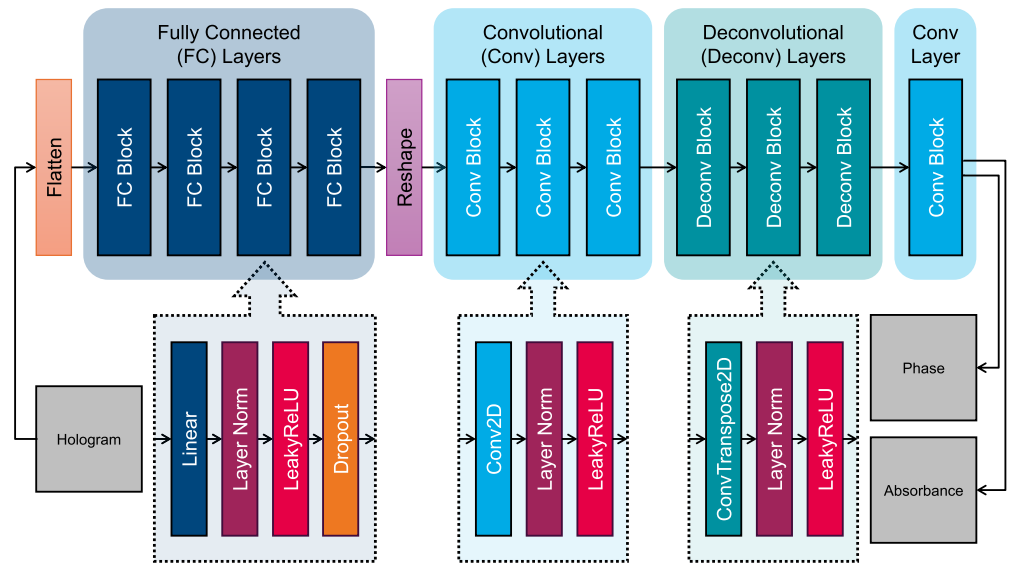

**Fig. 3.** The generator network $G$ predicts phase and absorbance variables from the input hologram (i.e., the measured intensity) and consists of fully connected, convolutional, and deconvolutional layers, and a final convolution.

As an alternative to the generator based on fully connected layers, a U-Net-based generator—specifically designed for phase retrieval tasks—can also be employed [63,64]. Both generator types demonstrate the capability to reconstruct high-quality data for phase retrieval applications. However, their performance varies across different scenarios and parameter combinations.

The discriminator employs a standard CNN classifier configuration, composed of multiple convolutional layers that feature dropout and utilize various strides and kernel sizes. The LeakyReLU activation function is used in all layers. A dense layer completes the discriminator, transforming the flattened output of the preceding layer into a scalar value in [0,1]. Default parameters are specified for both the generator and the discriminator, which offers a starting point for the model configuration. These parameters can be modified to fine-tune the model's performance across various data scenarios.

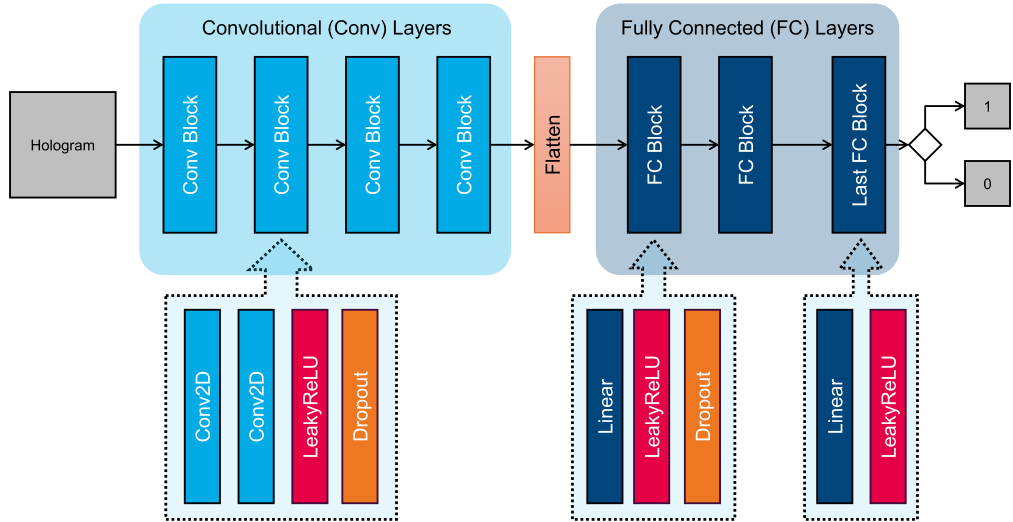

**Fig. 4.** The discriminator network *D* starts with convolutional layers and then pass through fully connected layers to classify whether the reconstructed hologram is a true hologram.

**Implementation and optimization** For the implementation of SelfPhish using TensorFlow, we apply three different normalizations, $T_1$, $T_2$, and $T_3$, to the (flat-field corrected) input hologram, the generated phase, and the generated absorbance, respectively, with

$$T_1(X) = \frac{X - \overline{X}}{\text{std}(X)}, \quad T_2(X) = \frac{T_1(X)}{\max[T_1(X)]}, \quad \text{and} \quad T_3(X) = \alpha\left[1 - T_2(X)\right], \tag{16}$$

where std($X$), max($X$) and $\overline{X}$ denote standard deviation, maximum and mean of $X$, respectively. The absorption factor $\alpha$ determines the *average* strength of the attenuation. This normalization method imposes no requirement for the phase to be non-positive or the absorbance to be non-negative. Therefore, it allows the network to penalize positive or negative values of the phase and absorbance, respectively. For the implementation in PyTorch, we only rescale the generated absorbance by $\alpha$, and asserted a condition to render the absorbance non-negative and the phase non-positive. The conditions arise naturally because of the interaction of X-rays with matter. Without imposing this condition, we do not obtain quantitative results.

We observed that the choice of normalization depends on the framework used, with the selected normalization method performing well in our TensorFlow and PyTorch implementations. Selecting values close to the mean ratio of $\beta/\delta$ helps accelerate convergence during iteration, as it rescales the network's output to lie closer to the expected physical range. However, this does not imply a direct coupling between phase and absorbance, as both are still reconstructed independently through separate output channels. We also apply batch normalization to all convolution layers, as suggested in [65], which accelerates convergence and improves the accuracy of the final results in our tests. We utilize the Adam optimizer [66] to address the Min-Max objective function described in Eq. (14) for the TensorFlow and PyTorch implementations. Selected for its exemplary performance in our specific application, this optimizer delivers an excellent balance between rapid convergence and stability. A crucial determinant of both the convergence rate and the quality of the reconstruction is the learning rate. To this end, we have predefined default learning rates for both the generator and the discriminator as initial guidelines. However, these rates are flexible and can be adjusted to suit various data conditions, enhancing model performance in a variety of situations.

## 3. Results

We evaluated SelfPhish for both simulated data where the ground truth, i.e., phase and absorbance, are known (Sec. 3.1 and 3.2), and experimental data (Sec.3.3). For simulated data, we have analyzed the performance of the model with varying Fresnel numbers, which translates into varying distances of the detector at a fixed X-ray energy and pixel size (Sec. 3.1). We further evaluated the model for different Poisson and Gaussian noise levels (Sec. 3.2). Finally, we validated the performance of SelfPhish with experimental data of a magnesium alloy (Sec. 3.3.1) and a spider hair (Sec. 3.3.2) to show its ability to deal with real measurement conditions.

### 3.1. Evaluation of SelfPhish across varying Fresnel numbers

We generated a phantom object as ground truth to evaluate the phase retrieval with SelfPhish. The phantom consists of a varying number of randomly placed solid spheres, each with random positions and sizes, within a cubic volume of $512^3$ voxels. The projections of these spheres were calculated and used as phase maps by setting $\delta = 3.1 \times 10^{-6}$. For the absorbance, the phase-attenuation duality was assumed with $\beta/\delta = 0.001$. The resulting phase shift and absorbance were used as input for the forward propagation of the exit wave field to generate the hologram. Using an X-ray energy of $E = 15$ keV and a pixel size of $d = 100$ nm, the performance of SelfPhish was evaluated for different Fresnel numbers $F$ in the range of 0.0012 to 0.12 corresponding to propagation distances between $z = 0.001$ m and $z = 0.1$ m. The size of simulated holograms is $512 \times 512$ pixels. To propagate the wave field, the projection was padded to $1024 \times 1024$ pixels, ensuring sufficient sampling of the propagator. For retrieving the phase an absorption factor of $\alpha = 0.001$ and 2000 iterations were used. The reconstruction process was run with the TensorFlow backend. It takes approximately 1.5 min with a NVIDIA A100 GPU.

Figure 5 shows a three-dimensional rendering of the simulated object, the corresponding simulated phase and hologram, the reconstructed phase and a map of the structural similarity index (SSIM) between the simulated and reconstructed phase. Figure S2 of the Supplement 1 presents the phase and absorbance maps reconstructed using SelfPhish, along with the normalized mean square error (NMSE) maps comparing the SelfPhish-reconstructed phase to the ground truth. Additionally, a comparison with conventional, non-iterative phase retrieval employing the linearized TIE is included. Results for a simulation using a strong absorbance with $\beta/\delta = 0.1$ are presented in Fig. S3 of the Supplement 1. For a Fresnel number of $F = 0.12$, the reconstructed phase is visually very similar to the simulated phase. To estimate the quality of the reconstruction with respect to the simulated phases, we employ the structural similarity index, where a value of 1 indicates a perfect match and 0 a complete mismatch. We observe that the interiors of the spheres exhibit a perfect match. However, the SSIM value decreases slightly around the edges of the spheres. Although significant local decreases in the SSIM value are observed, the reconstructed and simulated phases are visually very similar in these regions.

Lowering the Fresnel number by increasing the propagation distance leads to the formation of more fringes in the simulated hologram, initially resulting in an enhanced quality of the phase reconstruction. These results are illustrated in Fig. 6, which shows the simulated holograms and the corresponding SSIM maps between the reconstructed and simulated phase. At $F = 0.012$, the quality of the phase reconstruction reaches its maximum as shown by the SSIM map indicating an almost perfect match, except for minor discrepancies in areas where the projected edges of multiple spheres intersect. Below $F<0.01$, the quality of reconstruction begins to diminish and at $F = 0.001$, the quality deteriorates significantly. Here, the decline in the SSIM value is observed not just around but also within the projected spheres. This optimum with respect to the Fresnel number can be attributed to the fact that as the Fresnel number decreases and approaches the Fraunhofer diffraction regime, the diffraction fringes become larger and wider, eventually extending beyond the detector's field of view. In the absence of field-of-view limitations, the reconstruction is likely to continue to improve as the hologram captures increasingly more propagation-based information, i.e., diffraction fringes, in contrast to the hologram at $z = 0$, which contains only attenuation contrast and no phase information.

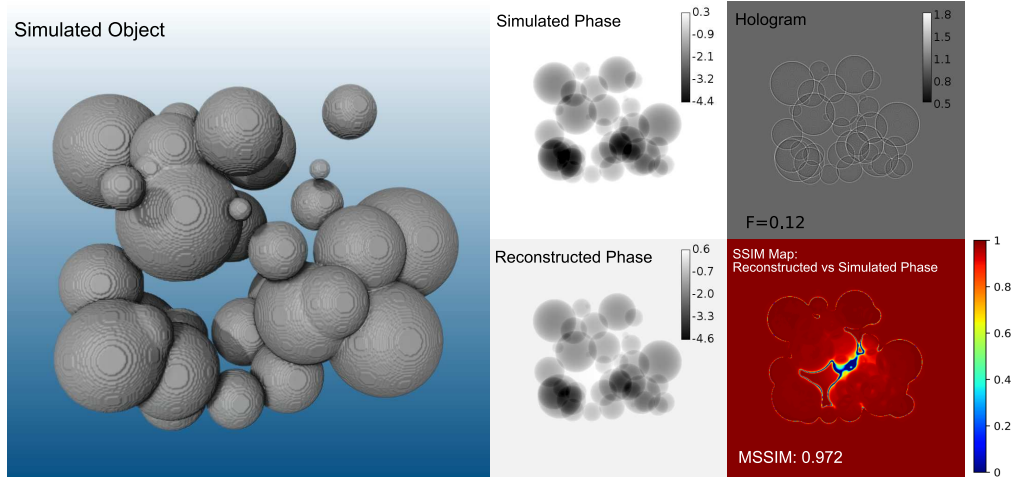

**Fig. 5.** Simulation of a 3D phantom composed of differently sized spheres, along with corresponding phase maps and hologram. Left image: 3D rendering of the simulated object. Middle column: Corresponding simulated and reconstructed phase maps. Top right image: Simulated input hologram. Bottom right image: Map of the structural similarity index measure (SSIM) between simulated and reconstructed phase with a mean SSIM of *MSSIM* = 0.971. A Fresnel number of 0.12 was used corresponding to an X-ray energy of $E = 15$ keV, a pixel size of $d = 100$ nm, and a propagation distances of $z = 0.001$ m.

### 3.2. Evaluations of SelfPhish under varying noise

Noise is another major factor affecting the quality of the retrieved phase maps. To assess its impact, we evaluated the performance of SelfPhish using different levels of Gaussian and Poisson noise. A realistic test object was simulated using data from a microtomography experiment of a biodegradable magnesium-gadolinium alloy (Mg-10Gd) screw implanted in cortical bone. This simulated object was used to calculate the phase and absorbance, and the resulting wave field was forward propagated using an energy of 50 keV, a pixel size of 0.64 µm, and a propagation distance of 1.86 µm, corresponding to a Fresnel number $F = 0.008$. The upper row of Fig. 7 shows a rendering of the simulated object, the corresponding simulated phase map, and the hologram without noise.

The additive Gaussian noise $\mathcal{N}$ follows a normal distribution for the intensity $I$ with $I_{\mu,\sigma} = I + \mathcal{N}(\mu, \sigma)$, zero mean $\mu = 0$, and a standard deviation of $\sigma = 0.1, 0.2, 0.3$ and $0.4$. The non-additive Poisson noise follows a conditional probability function $P_I$. To test different noise levels, we used a scaling factor $\kappa$ with $P_I(\kappa) = \kappa P(\frac{1}{\kappa} I)$ and $\kappa = 0.001, 0.01, 0.1, 0.5$ and $1$.

For the results presented in this section, 2000 iterations were used which completed in 325 s. We have observed that, as the noise gets stronger, the model starts to learn the noise of the intensity in the later stages of the learning and the quality of the reconstruction decreases. Therefore, early stopping is important in real life experiments. The two bottom rows of Fig. 7 display the simulated noisy holograms along with corresponding phase reconstructions for a single noise level. To quantitatively assess the reconstruction quality, we display the structural similarity index measure (SSIM) maps, along with corresponding mean structural similarity index measure (MSSIM) and peak signal-to-noise ratio (PSNR) values, in Fig. 8. These metrics were calculated from reconstructed phase images obtained from holograms with varying noise levels (Gaussian and Poisson). The retrieved phase maps along with the NMSE maps for Gaussian and Poisson noise are presented in Figs. S4 and S5 of the Supplement 1, respectively. Additionally, for Gaussian noise, a comparison with conventional, non-iterative phase retrieval employing the linearized TIE is included in Fig. S4. As demonstrated in Fig. 8, SelfPhish effectively recovers the phase from noisy holograms under low to moderate noise conditions. Although higher noise levels (e.g., $\kappa \geq 0.5$ and $\sigma \geq 0.5$) pose challenges and can lead to a loss of finer details, SelfPhish still maintains acceptable accuracy. Overall, these results confirm SelfPhish's robustness and reliability, making it a promising method for practical phase retrieval applications even in noisy experimental conditions.

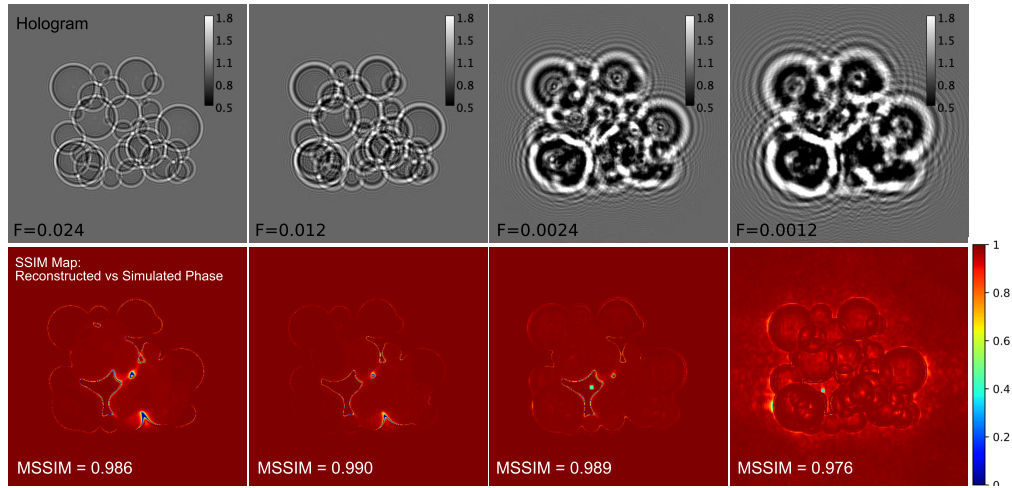

**Fig. 6.** Simulated holograms and SSIM maps of the corresponding reconstructions for different Fresnel numbers. Holograms were simulated using an X-ray energy of $E = 15$ keV, a pixel size of $d = 100$ nm, and propagation distances of $z = 0.005$m, $0.010$m, $0.050$m and $0.100$ m corresponding to Fresnel numbers of $F = 0.024, 0.012, 0.0024$ and $0012$, respectively. Please note that the colormap for all holograms is constrained to the same range of values.

### 3.3. Experimental validation of SelfPhish

We performed in-line X-ray holography measurements on a corroded biodegradable magnesium alloy [67] and a spider attachment hair [68] at the P05 imaging beamline (IBL) [69], operated by Helmholtz-Zentrum Hereon at the synchrotron-radiation source PETRA III (Deutsches Elektronen-Synchrotron DESY, Hamburg, Germany). The experiments were conducted in the holographic regime, characterized by small Fresnel numbers ($F \ll 1$), but not reaching the Fraunhofer regime. Detailed experimental parameters, including the effective propagation distance and effective pixel size — computed using the Fresnel scaling theorem [5] — are summarized in Table 1. For the magnesium alloy sample, we also show the retrieved absorbance along with the phase information.

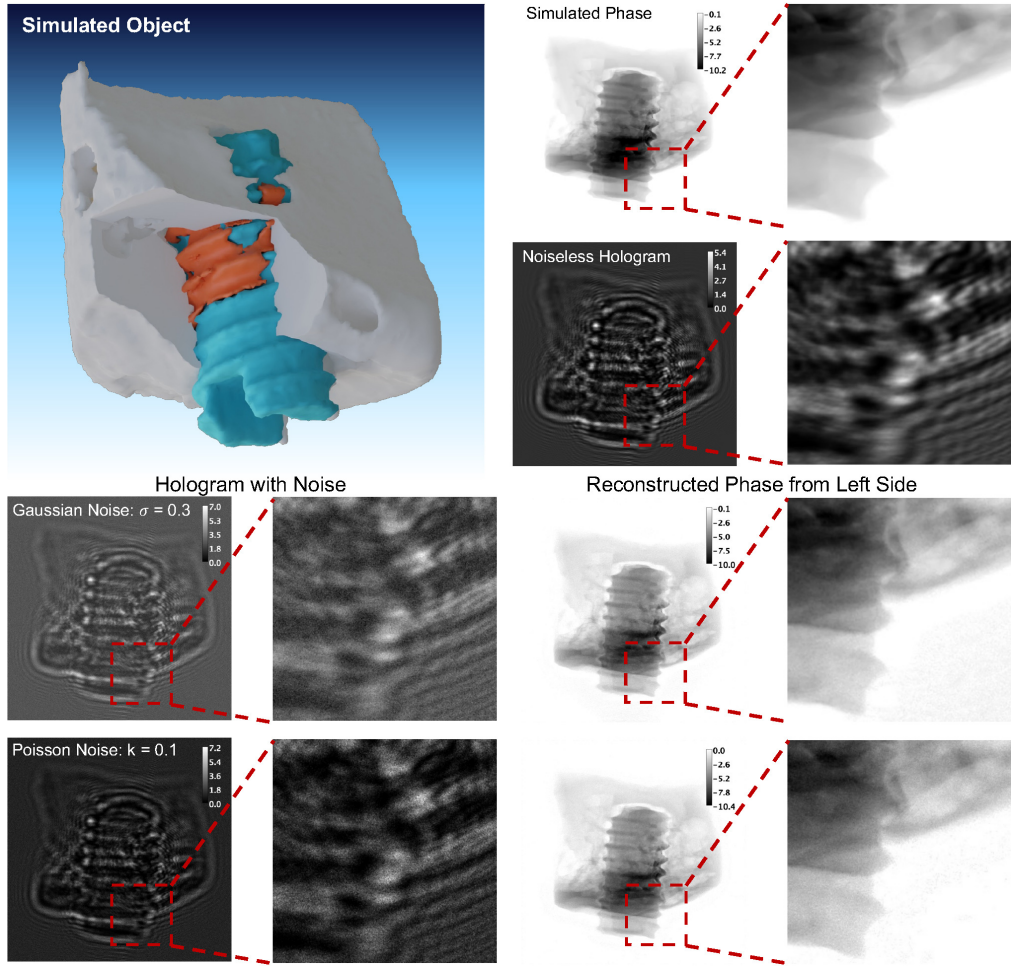

**Fig. 7.** Simulation of a biodegradable screw implant in bone and phase retrieval for noisy data. Top left: Rendering of the simulated object made up of cortical bone (gray), a magnesium-gadolinium (Mg-10Gd) screw (blue), and a corrosion layer (red). Top right: Corresponding simulated phase and hologram without noise. Bottom left: Simulated holograms with Gaussian and Poisson noise, respectively. Bottom right: Phase reconstructions using SelfPhish corresponding to the holograms on the left. The red dashed box and lines indicate zoomed-in regions.

#### 3.3.1. Corroded biodegradable magnesium alloy

The sample shown in Fig. 9 is a corroded biodegradable magnesium-based alloy [67], which presents significant challenges for phase retrieval due to large phase shifts and incomplete background correction. Data were collected as part of the SmartPhase project [70]. This example highlights SelfPhish's ability to robustly and quantitatively retrieve both phase and absorbance, demonstrating its strength to address complex phase retrieval scenarios.

We introduce prior knowledge into the model by imposing non-negativity for the absorbance using the ReLU function and non-positivity for the phase using −ReLU. This ensures that the reconstructed absorbance and phase remain physically meaningful. Experimental parameters, including the absorption factor $\alpha$) are detailed in Table 1. Phase retrieval using an input hologram with $1024 \times 1024$ pixels takes from 2.8 to 3.2 min for 2000 iterations depending on the padding of the complex wave field. During analysis, we observed that imperfect flat-field correction during the pre-processing of the hologram resulted in a non-zero background in the phase. Interestingly, this issue is absent in the absorbance data, suggesting that the phase is more sensitive to inaccuracies in flat-field correction than the absorbance. This behavior is similar to the well-known observation of large-scale or cloud-like distortions in classical TIE- or CTF-based phase retrieval, which are typically due to residual non-homogeneous attenuation, imperfect flat-field correction or noise. The color bar highlights the presence of a strong phase shift, which presents significant challenges for most phase retrieval algorithms. While SelfPhish demonstrates superior performance in reconstructing the absorbance compared to [62], the latter method shows an advantage in achieving better background reconstruction for the phase.

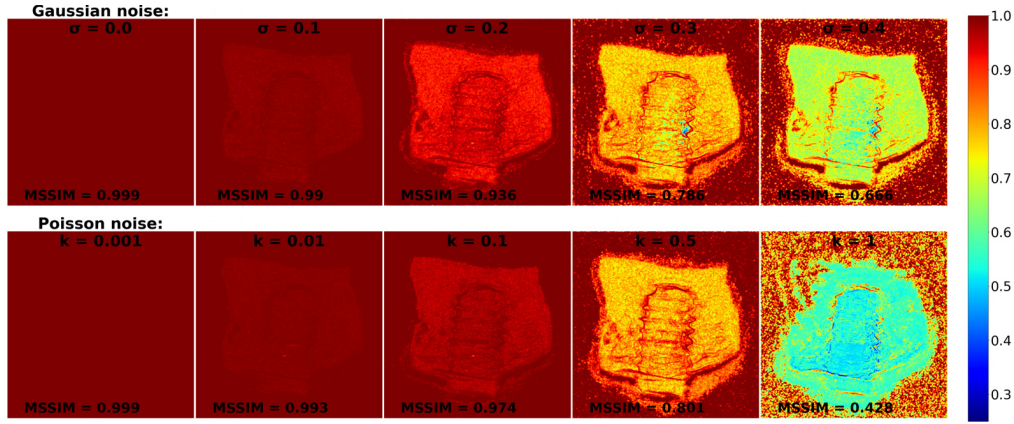

**Fig. 8.** SSIM maps between the simulated and reconstructed phase for holograms with Gaussian and Poisson noise and increasing noise level from left to right. Top row: Gaussian noise with standard deviation $\sigma$. Bottom row: Poisson noise with noise factor $\kappa$. The mean SSIM value (MSSIM) is shown at the bottom of each cell. The PSNR values between the simulated and reconstructed phases from left to right are 56.23, 47.33, 36.21, 28.13, 24.87 for Gaussian noise (top row) and 54.11, 46.21, 42.03, 30.57, 15.24 for Poisson noise (bottom row).

**Table 1. Experimental and reconstruction parameters for the in-line X-ray holography measurement of the corroded biodegradable magnesium alloy and the spider hair attachment.**

| Parameters | Magnesium alloy | Spider hair |
| --- | --- | --- |
| Energy | 11 keV | 11 keV |
| Wavelength | 0.11 nm | 0.11 nm |
| Magnification | 36.8 | 248 |
| Focus-object distance | 470.5 mm | 79.95 mm |
| Sample-detector distance | 19.2 m | 19.6 m |
| Detector pixel size | 6.5 µm | 6.5 µm |
| Effective distance | 521.5 mm | 78.9 mm |
| Effective pixel size | 176.5 nm | 26.2 nm |
| Fresnel number rescaled | $5.32 \times 10^{-4}$ | $7.72 \times 10^{-5}$ |
| Absorption factor ($\alpha$) | $5 \times 10^{-4}$ | $1 \times 10^{-3}$ |

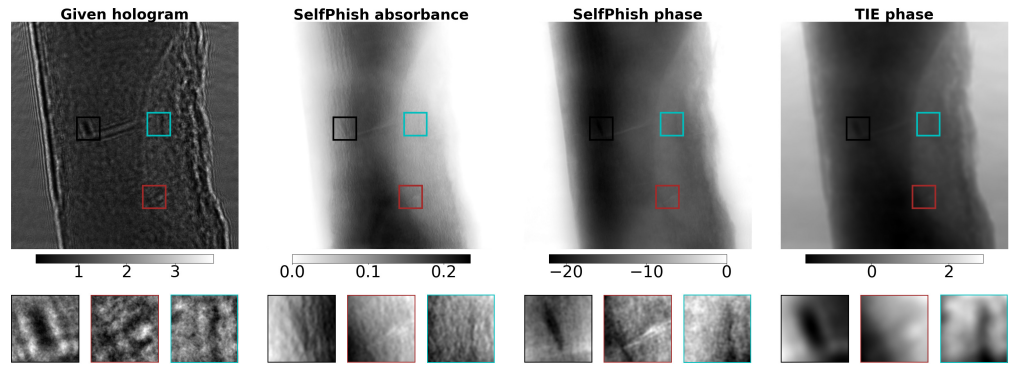

**Fig. 9.** Hologram from a nano-tomography experiment of a corroded magnesium alloy (left). Phase (second column) and absorbance (third column) retrieved using SelfPhish for an object exhibiting strong phase shifts, as indicated by the color bar. SelfPhish yields sharper, more detailed phase and absorbance maps than the conventional, non-iterative phase retrieval (right) employing the linearized transport of intensity equation (TIE). The zoom boxes show regions with strong propagation-based phase contrast fringes in the hologram, along with the corresponding phase and absorbance, which were well retrieved by SelfPhish.

#### 3.3.2. Spider attachment hair

We conducted in-line holography measurements on a spider hair sample, collecting a complete set of 180 holograms evenly spaced over 180°. Each hologram was processed with phase retrieval using SelfPhish, with the absorption factor $\alpha$ set to 0.001. The retrieval involved 700 iterations per projection, taking approximately 98 s each. The phase reconstruction, shown in Fig. 10, consistently captured fine biological details at all projection angles, clearly highlighting small spine structures, touch-sensitive hairs (*microtrichia*), and spatulae. Additionally, we present the phase reconstruction using the linearized TIE method. This comparison clearly highlights the need for more advanced phase retrieval techniques, such as SelfPhish.

Using these reconstructed phases, we performed a 3D tomographic reconstruction of the spider hair using TomoPy [71]. The resulting tomogram was visualized in 3D using Avizo (Thermo Fisher Scientific). The consistent quality of the retrieved phase at various angles ensured reliable 3D imaging of the sample, capturing detailed biological features with clarity comparable to those reported by J. Dora et al. [62]. These results demonstrate SelfPhish's practicality and effectiveness for 3D imaging of biological samples using in-line holography.

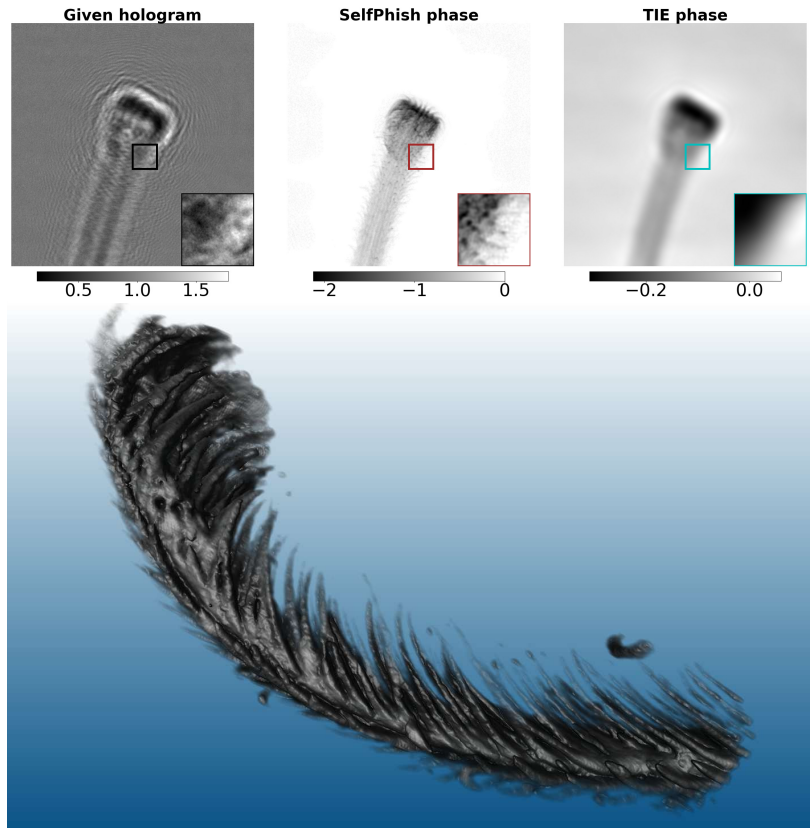

**Fig. 10.** In-line X-ray hologram of a spider hair sample and corresponding phase reconstructions. Top row: Flat-field corrected input hologram (left), phase map reconstructed with SelfPhish (center), and phase map obtained using the linearized transport of intensity equation (TIE) (right). Bottom: 3D visualization of the reconstructed spider hair.

## 4. Discussion and conclusion

Phase retrieval using only a single-distance intensity measurement represents a classic example of an ill-posed inverse problem, where a perfect solution is not guaranteed for any solver. As discussed in the Introduction, deep image prior approaches can suffer from optimization getting stuck in local minima [55,56]. SelfPhish mitigates this issue by incorporating a discriminator network, following the principles of the GANrec framework [59], which replaces the traditional loss function with an adversarial objective. This design choice reduces the likelihood of getting trapped in poor local minima, as supported by the ablation study presented in Sec. 1 of the Supplement 1, where the discriminator was removed from the network (see Fig. S1, as well as Figs. S4 and S5). SelfPhish demonstrates stable performance and high-quality reconstructions for both simulated and experimental data. In scenarios involving very high noise levels (or inconsistent data resulting from, e.g., beam instabilities during measurement), its performance becomes less stable, with convergence issues arising occasionally. In addition, network overfitting may occur when the reconstruction process exceeds a certain number of iterations.

For the reconstruction, either randomly initialized weights or pre-trained weights can be used. Random weights enable the algorithm to search for the whole solution space, resulting, however, in long reconstruction times. When reconstructing images with similar features, e.g.

sequences of projections from a tomography experiment, high-quality reconstructions can be performed utilizing the trained weights from the prior reconstruction of previous projections. This methodology has been used, for example, in the reconstruction of the tomographic data set of the spider hair. Likewise, high-resolution data can be handled. Using a hierarchical approach, the results from a downsampled data set can be used to initialize the network for the reconstruction on a finer highly-resolving grid.

The number of iterations, the absorption factor $\alpha$, the coefficient of the data loss term $\gamma$, and the learning rate of the networks can differ between various applications and may require further calibration to achieve optimal performance. An initial estimate of the absorbance factor can be derived from the average absorbance of the flat-field-corrected hologram. In addition, we are investigating methods to determine these parameters automatically.

Deep neural networks have proven to be effective in learning from physics-based models and solving inverse problems with high accuracy. The application of SelfPhish for phase retrieval demonstrates its flexibility as an inverse solver, capable of handling a variety of ill-posed problems when the forward model is well defined. In X-ray holography reconstruction, SelfPhish successfully retrieves both phase and absorbance from single-distance intensity measurements, with its performance confirmed through simulations and experimental data. Unlike traditional algebraic approaches, it does not rely on approximations of image formation physics or require extensive fine-tuning of boundary conditions in iterative phase retrieval methods. By directly mapping the measured intensity to the underlying phase, SelfPhish offers a practical and efficient solution to the phase retrieval problem.

In future investigations, we intend to extend its applicability to more complex phase retrieval scenarios and enhance its performance. This includes the adaption of SelfPhish for mainly attenuating samples with edge-enhancement fringes only, the utilization of the 3D information of the tomographic data sets, or the incorporation of partially coherent beam properties in the forward model. In particular, we will evaluate the reconstructed absorbance, which is often not available or of minor quality, using other approaches.

The successful application of the SelfPhish architecture to both tomography and phase retrieval demonstrates its versatility in tackling fundamentally distinct challenges. For tomography, the network learns the inverse Radon transform. For phase retrieval, the model enforces physical properties by using the Fresnel propagator. We hypothesize that nonphysical solutions, which are admissible because of the ill-posedness of the problem, are mitigated due to the properties of GANs.

**Funding.** Helmholtz Association (ZT-I-PF-5-); Deutsche Forschungsgemeinschaft (192346071,446 SFB 986, project Z2); Bundesministerium für Bildung und Forschung (031L0202A, 05D23CG1); U.S. Department of Energy (BR #KC0406021).

**Acknowledgment.** Parts of this research were supported by the BMBF project *Multi-task Deep Learning for Large-scale Multimodal Biomedical Image Analysis (MDLMA)* (BMBF project number 031L0202A), the Helmholtz AI project *Universal Segmentation Framework (UniSeF)*, and the Hereon project *Holistic Data Analysis (HoliDAy)* of the Innovation-, Information-& Biologisation-Fonds (I²B). The authors gratefully acknowledge financial support from the Deutsche Forschungsgemeinschaft (DFG) (project No. 192346071, SFB 986, project Z2). Parts of this research were supported by the BMBF Project *Ein KI-basiertes Framework für die Visualisierung und Auswertung der massiven Datenmengen der 4D-Tomographie für Endanwender von Beamlines* (KI4D4E) (BMBF project number 05D23CG1). This research also used resources from the DOE project *Intelligent Acquisition and Reconstruction for Hyper-Spectral Tomography Systems (HyperCT, project No. BR #KC0406021)*. We acknowledge the Deutsches Elektronen-Synchrotron DESY (Hamburg, Germany), a member of the Helmholtz Association HGF, for the provision of beamtime, related to the proposal I-20180109 and I-20191467 at the P05 imaging beamline (IBL) at PETRA III at DESY. This research was supported in part through the Maxwell computational resources operated at DESY. The research leading to this result has been supported by Hi-Acts, an innovation platform under the grant of the Helmholtz Association HGF.

**Disclosures.** The authors declare no conflict of interest.

**Data availability.** The source code and a subset of the data used in this study are publicly available [72]. PyTorch implementation has been archived on Zenodo [73].

**Supplemental document.** See Supplement 1 for supporting content.

# Self-supervised physics-informed generative networks for phase retrieval from a single X-ray hologram: supplement

Xiaogang Yang,[1,†,*] Dawit Hailu,[2] Vojtěch Kulvait,[2] Thomas Jentschke,[2] Silja Flenner,[2] Imke Greving,[2] Stuart I. Campbell,[1] Johannes Hagemann,[3] Christian G. Schroer,[3,4,5] Tak Ming Wong,[2,6] and Julian Moosmann[2,†]

[1]*NSLS-II, Brookhaven National Laboratory, Upton, NY, 11973, USA*
[2]*Institute of Materials Physics, Helmholtz-Zentrum Hereon, Max-Planck-Straße 1, 21502 Geesthacht, Germany*
[3]*Center for X-ray and Nano Science CXNS, Deutsches Elektronen-Synchrotron DESY, Notkestraße 85, 22607 Hamburg, Germany*
[4]*Department of Physics, Universität Hamburg, Luruper Chaussee 149, 22761 Hamburg, Germany*
[5]*Helmholtz Imaging, Deutsches Elektronen-Synchrotron DESY, Notkestraße 85, 22607 Hamburg, Germany*
[6]*Institute of Metallic Biomaterials, Helmholtz-Zentrum Hereon, Max-Planck-Straße 1, 21502 Geesthacht, Germany*
[†]*The authors contributed equally to this work.*
[*]*yangxg@bnl.gov*

# Self-supervised physics-informed generative networks for phase retrieval from a single X-ray hologram: supplemental document

## 1. ABLATION STUDY ON THE DISCRIMINATOR NETWORK

To evaluate the role of the discriminator during phase retrieval, we compared our adversarial architecture with a baseline model that uses the same generator but omits the discriminator, resembling a deep image prior (DIP) setup. For this evaluation, we used the bone implant simulation from Sec. 3.2. Figure S1 shows the $L_1$ distances over 2000 model iterations: one between the input (simulated intensity) $I$ and the generated hologram $I^*$ and the other between the ground truth phase $\phi$ and the reconstructed phase $\phi^*$. While the intensity difference (i.e. the discrepancy minimized during the network training) remains similar for both models, the phase error is consistently lower when the discriminator is present. This indicates that the model with a discriminator reconstructs a phase that is more faithful to the ground truth, even though neither model is directly optimized to minimize this $L_1$ phase error.

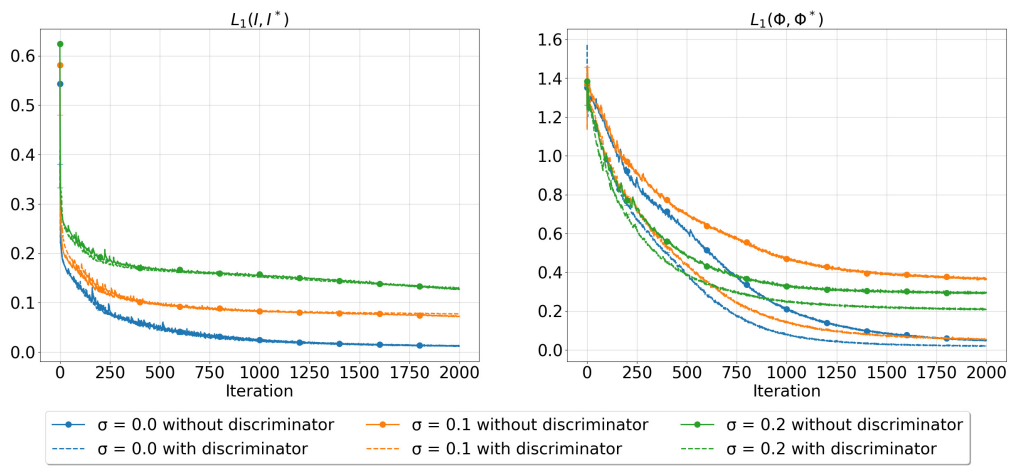

**Fig. S1.** Plots of the $L_1$ distance of the intensity and phase for models with and without a discriminator and three noise levels. Left: $L_1$ distance between input (simulated) $I$ and generated hologram $I^*$. Right: $L_1$ distance between ground truth, $\phi$, and reconstructed phase $\phi^*$. The discriminator-guided model achieves a lower phase error, demonstrating improved reconstruction.

To further assess the impact of the discriminator on phase retrieval, we evaluated the reconstructed phase across a range of noise levels using the normalized mean square error (NMSE) as a quantitative metric. the fourth and fifth row of Figs. S4 and S5, the GAN-based model consistently outperforms the generator-only baseline, achieving lower NMSE values across all tested noise conditions - except under very high Poisson noise levels, where accurate phase retrieval is generally not expected. The corresponding NMSE maps visually confirm that the discriminator-guided reconstructions are closer to the ground truth. This supports our hypothesis that the proposed adversarial architecture encourages solutions that are not only consistent with the forward model but also structurally faithful to the underlying phase distribution. Importantly, this improvement is achieved without requiring any paired training data, further highlighting the value of incorporating an untrained discriminator into self-supervised reconstruction frameworks.

In addition, Figs. S4 and S5 show the simulated holograms and the corresponding phases reconstructed using a model without and with discriminator, respectively. Phase retrieval required 325 s with the discriminator and 296 s without for 2000 iterations.

## 2. EVALUATION OF SELFPHISH ACROSS VARYING FRESNEL NUMBERS: ADDITIONAL INFORMATION

In Fig. S2 and S3, we present additional information for the solid sphere simulation for varying Fresnel numbers for weak ($\beta/\delta = 0.001$) and strong ($\beta/\delta = 0.1$) absorbances, respectively. The corresponding top row shows the simulated holograms. The corresponding second and third row shows the absorbance and phase retrieved by SelfPhish, respectively. The corresponding fourth row shows the phase retrieved using the linearized transport of intensity equation (TIE). For details about the TIE-based phase retrieval, see Sec. 4. The corresponding bottom row shows the NMSE map between ground truth phase and phase retrieved by SelfPhish.

As expected, the absorbance maps retrieved by SelfPhish are considerably worse in Fig. S2 due to the much weaker attenuation signal compared to Fig. S3. While the phase retrieval using SelfPhish in Fig. S2 shows consistent results across different Fresnel numbers, the TIE phase deteriorates considerably with increasing Fresnel number. For strong absorbance, SelfPhish performs worse than TIE for large Fresnel numbers, but better than TIE for small Fresnel numbers. This behavior is expected in cases of strong absorbances and large Fresnel numbers, where the contribution of propagation-based phase contrast to the intensity signal is significantly weaker than that of attenuation contrast.

## 3. EVALUATIONS OF SELFPHISH UNDER VARYING NOISE: ADDITIONAL INFORMATION

In Figs. S4 and S5, we present additional information for the biodegradable bone implant for Gaussian and Poisson noise, respectively. The corresponding top row shows the simulated holograms. The corresponding second and third row show the phase maps retrieved using a model without and with discriminator, respectively. The corresponding bottom two rows show the NSME maps between ground truth and phase retrieved with and without discriminator. In the bottom row of Fig. S4, we present phase maps retrieved using the linearized TIE. TIE-based phase retrieval consistently underperforms compared to SelfPhish across all Gaussian noise levels.

## 4. CONVENTIONAL, NON-ITERATIVE PHASE RETRIEVAL

For conventional, non-iterative phase retrieval, we employed the linearized transport of intensity equation (TIE) [1, 2] and assuming phase-attenuation duality [3, 4]. Here, the phase-attenuation duality ratio $\beta/\delta$ acts a regularization parameter and is empirically determined, typically selected to yield an approximately flat reconstructed background. At higher values of $\beta/\delta$, propagation induced phase-contrast fringes remain prominent in the retrieved phase. In contrast, lower values lead to increasingly blurred phase reconstructions. Table S1 summarizes the $\beta/\delta$ ratios employed for phase retrieval using the TIE method.

| Sample | $\beta/\delta$ |
|---|---|
| Solid spheres: weak absorbance (Sec. 3.1) | $6 \times 10^{-4}$ for all $F$ |
| Solid spheres: strong absorbance | 0.5 for all $F$ |
| Bone implant: $\kappa = 0.001, 0.01, 0.1, 0.5, 1$ (Sec. 3.2) | $0.006, 0.006, 0.004, 2.5 \times 10^{-3}, 6 \times 10^{-6}$ |
| Magnesium alloy (Sec. 3.3.1) | 10 |
| Spider hair (Sec. 3.3.2) | 0.04 |

**Table S1.** Phase-attenuation duality ratio $\beta/\delta$ used for the TIE-based phase retrieval.

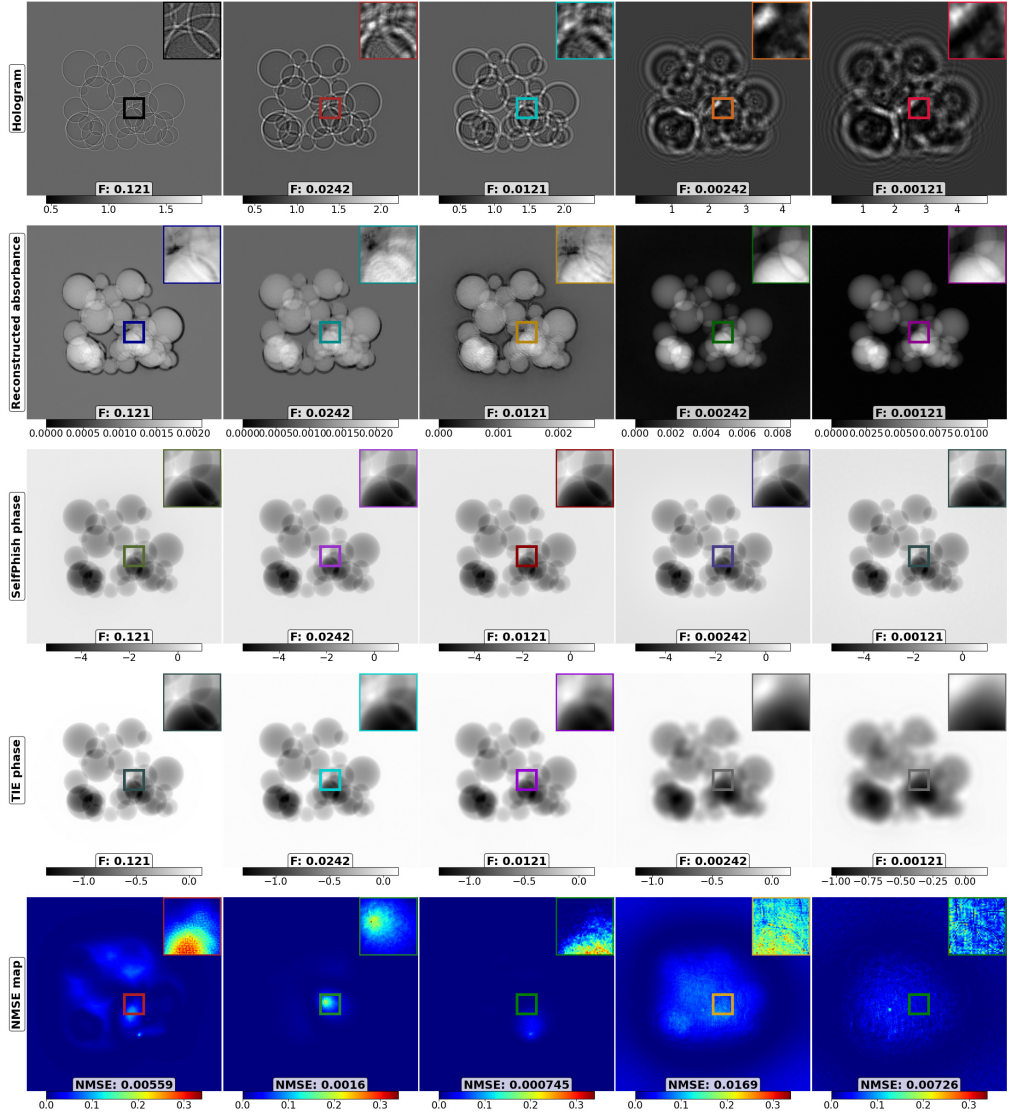

**Fig. S2.** A weak absorbance ($\beta/\delta = 0.001$) is used for simulating multiple holograms across multiple Fresnel numbers (top row). Using SelfPhish, we reconstructed the absorbance (second row) and phase (third row). For comparison, we show the phase reconstructed using the linearized TIE (fourth row). NMSE maps between ground truth and SelfPhish phase are shown in the bottom row.

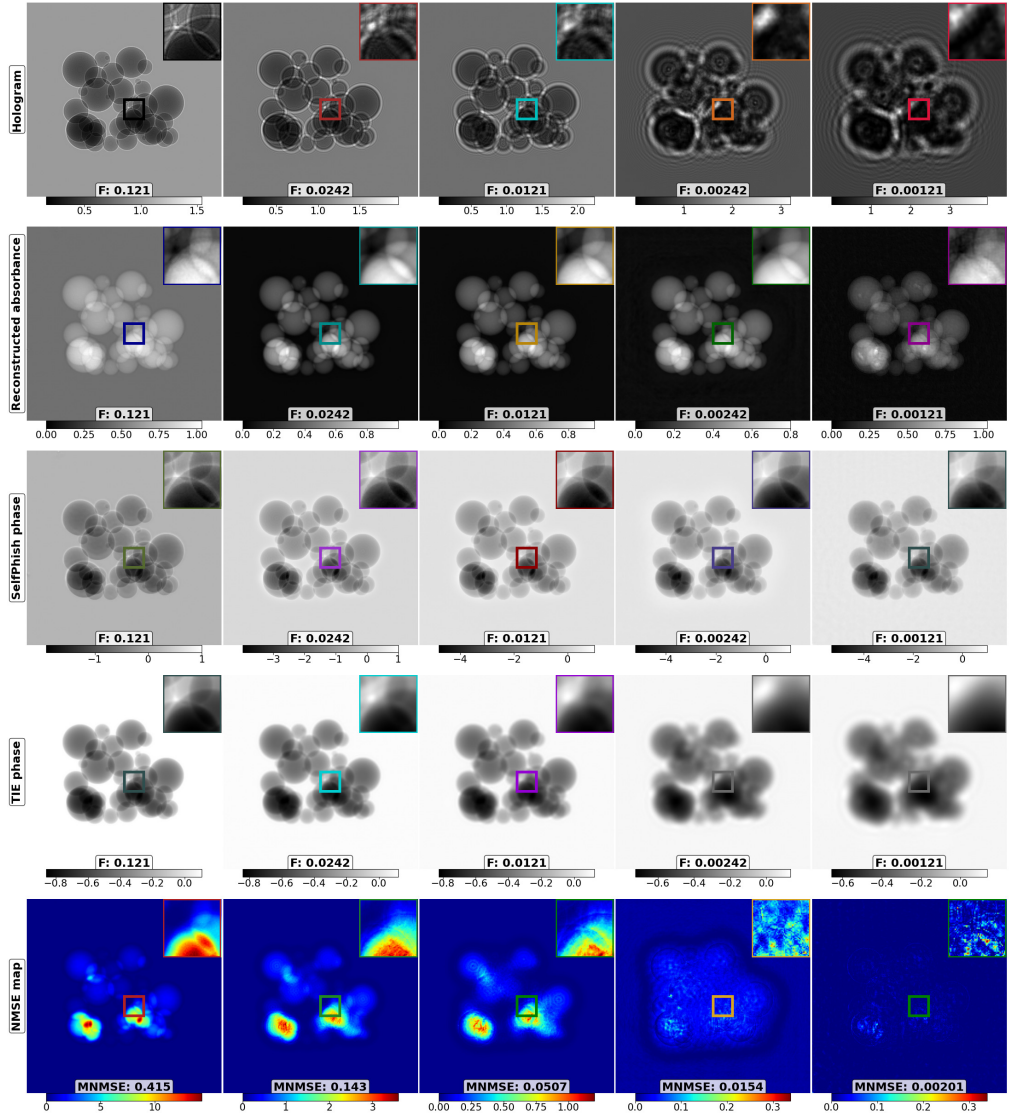

**Fig. S3.** A strong absorbance ($\beta/\delta = 0.1$) is used for simulating multiple holograms across multiple Fresnel numbers (top row). Using SelfPhish, we reconstructed the absorbance (second row) and phase (third row). For comparison, we show the phase reconstructed using the linearized TIE (fourth row). NMSE maps between ground truth and SelfPhish phase are shown in the bottom row.

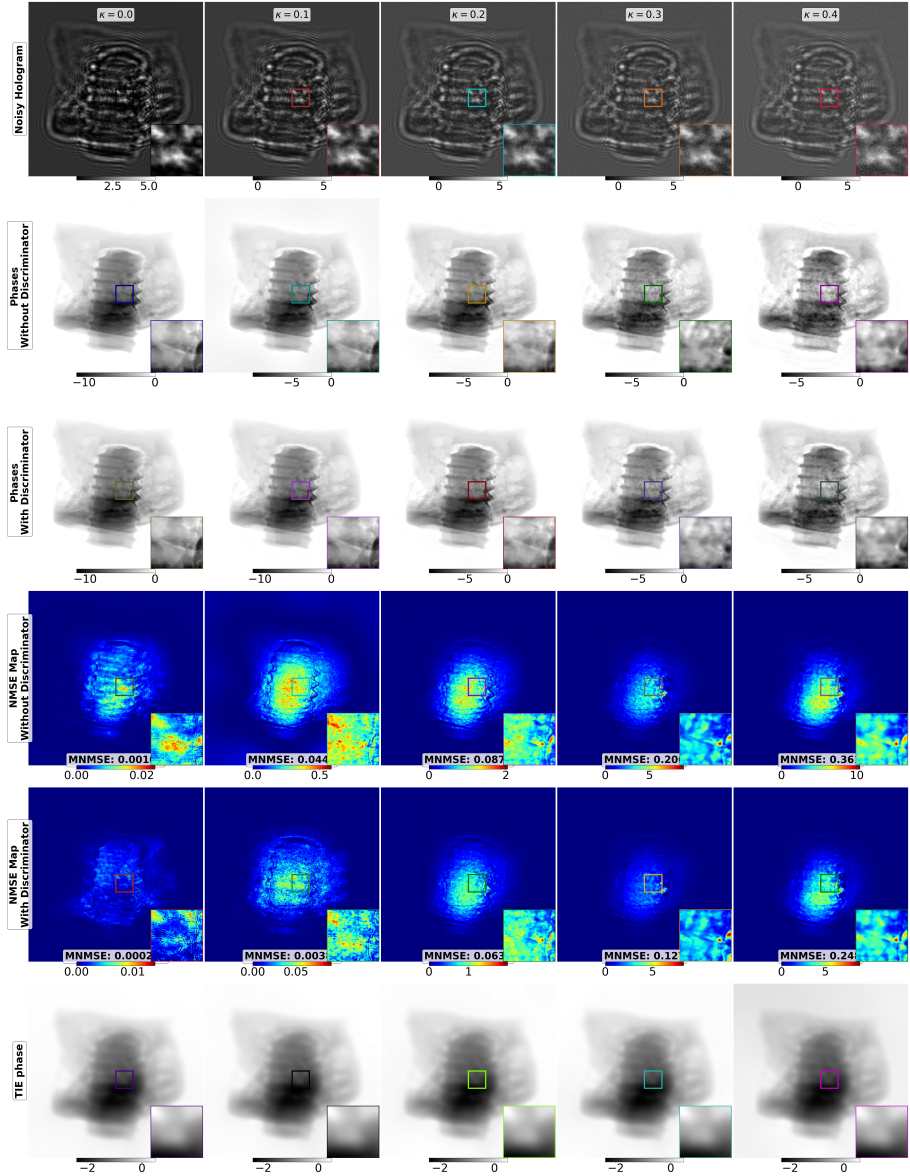

**Fig. S4.** Comparison of phases reconstructed from holograms with varying levels of Gaussian noise ($\sigma$) and for a model without and with discriminator. The respective phase maps (second and third row) are shown along with the normalized mean square error (NMSE) maps (fourth and fifth row) and corresponding mean NMSE values (MNMSE) between the simulated ground truth and reconstructed phase. Phase maps retrieved using the linearized TIE are shown in the bottom row. Please note that the colormap is adjusted to the range of gray values of each image individually to ensure optimal contrast.

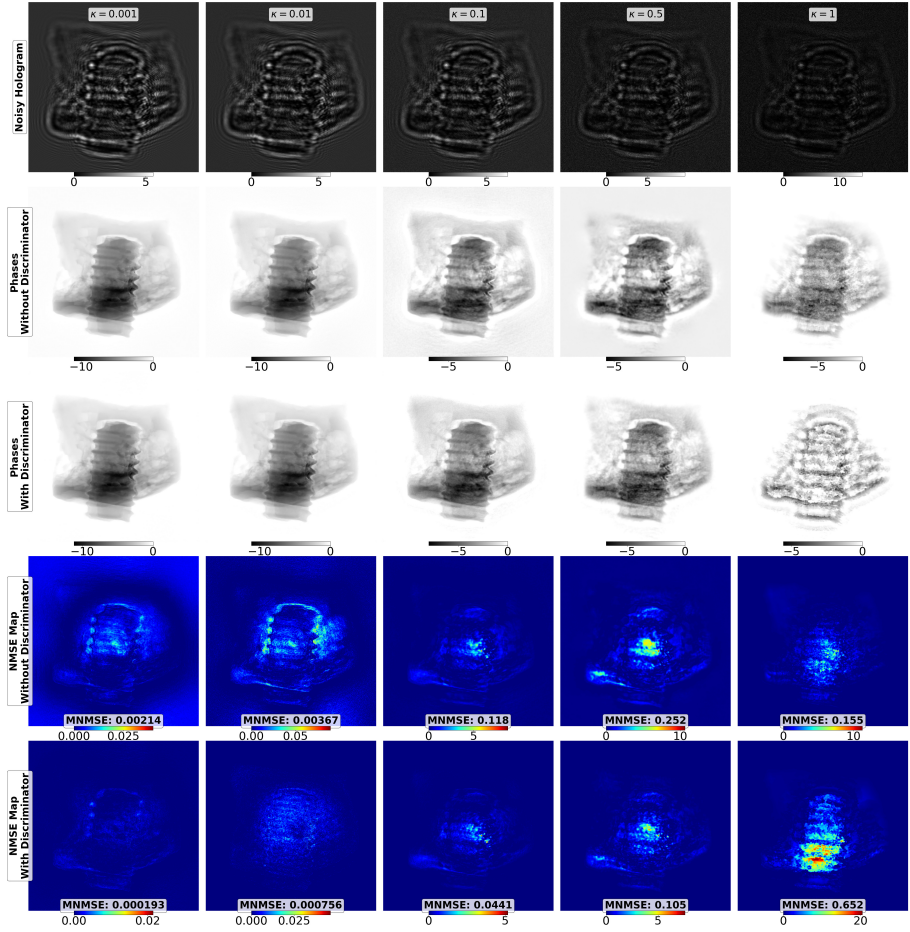

**Fig. S5.** Comparison of phases reconstructed from holograms (top row) with varying levels of Poisson noise ($\kappa$) and for a model without and with discriminator. The respective phase maps (second and third row) are shown along with the normalized mean square error (NMSE) maps (fourth and fifth row) and corresponding mean NMSE values (MNMSE) between the simulated ground truth and reconstructed phase. Please note that the colormap is adjusted to the range of gray values of each image individually to ensure optimal contrast.